\documentclass{aa}
\usepackage{graphicx}
\usepackage{amssymb}
\begin{document}
\title{The acoustic spectrum of $\alpha$~Cen~A\thanks{Based on observations
collected with the CORALIE echelle spectrograph on the 1.2-m Euler Swiss
telescope at La Silla Observatory, ESO Chile}}
\author{F. Bouchy and F. Carrier}
\offprints{F. Bouchy \\
           \email{Francois.Bouchy@obs.unige.ch}}
\institute{Observatoire de Gen\`eve, 51 ch. des Maillettes, 1290 Sauverny,
           Switzerland \\}
\date{Received 17 January 2002 / Accepted 7 May 2002}
\authorrunning{Bouchy \& Carrier}
\titlerunning{The acoustic spectrum of $\alpha$~Cen~A}
\abstract{
This paper presents the analysis of Doppler p-mode observations of the
G2V star $\alpha$~Cen~A obtained with the spectrograph CORALIE in May 2001.
Thirteen nights of observations have made it possible to collect 1850 radial
velocity measurements with a standard deviation of about 1.5 m\,s$^{-1}$.
Twenty-eight oscillation modes have been identified in the power spectrum
between 1.8 and 2.9~mHz with amplitudes in the range 12 to 44 cm\,s$^{-1}$.
The average large and small spacing are respectively equal to 105.5 and 5.6~$\mu$Hz. A comparison with stellar models of $\alpha$~Cen~A is presented.
}

\maketitle

\def\cms{\,cm\,s$^{-1}$}      
\def\ms{\,m\,s$^{-1}$}        
\def\kms{\,km\,s$^{-1}$}      
\def\vsini{$v$\,sin\,$i$}     
\def\m2s2{\,m$^{2}$\,s$^{-2}$} 

\section{Introduction}

The lack of observational constraints leads to serious uncertainties in the
modeling of stellar interiors and stellar evolution.
The measurement and characterization of oscillation modes in solar-like
stars is an ideal tool to test models of stellar inner structure and
theories of stellar evolution.

The discovery of propagating sound waves, also called p-modes, in the
Sun by Leighton et al. (\cite{leighton62}) and their interpretation by Ulrich
(\cite{ulrich70}), has opened a new area in stellar physics.
Frequency and amplitude of each oscillation mode depend on the physical
conditions prevailing in the layers crossed by the waves and provide a
powerful seismological tool.
Helioseismology led to major revisions in the ``standard model'' of the Sun
and provided for instance measures of the inner rotation of the Sun, the
size of the convective zone and the structure of the external layers.

Solar-like oscillation modes generate periodic motions of the stellar
surface with periods in the range 3 - 30 minutes but with extremely small
amplitudes. Essentially two methods exist to detect such a motion:
photometry and Doppler spectroscopy. In photometry, the oscillation 
amplitudes of solar-like stars
are in the range 2 - 30~ppm while they are in the range
10 - 150~{\cms} in radial velocity measurements.
Photometric measurements made from the ground are strongly limited by
scintillation noise. To reach the needed accuracy requires
observations made from space.
In contrast, Doppler ground-based measurements have recently shown their
capability to detect oscillation modes in solar-like stars.

The first good evidence of excess power due to mode oscillations was obtained
by Martic et al. (\cite{martic99}) on the F5 subgiant Procyon. Bedding et
al. (\cite{bedding01}) obtained a quite similar excess power for the G2 
subgiant $\beta$ Hyi. These two results have independently been confirmed by our
group based on observations made with the CORALIE spectrograph (Carrier et al.
\cite{carrier01a}, \cite{carrier01b}).

A primary target for the search for p-mode oscillations is the solar twin $\alpha$~Cen~A (HR5459).
Several groups have already made thorough attempts to detect the signature
of p-mode oscillations on this star. Two groups claimed mode detections
with amplitudes 3.2\,-\,6.4 greater than solar (Gelly et al. \cite{gelly86}; 
Pottasch et al. \cite{pottasch92}). However, this was refuted by three
other groups who obtained upper limits of mode amplitudes of 1.4\,-\,3 times solar (Brown \& Gilliland \cite{brown90}; Edmonds \& Cram \cite{edmonds95};
Kjeldsen et al. \cite{kjeldsen99}). More recently Schou \& Buzasi
(\cite{schou01}) made photometric observations of $\alpha$~Cen~A with
the WIRE spacecraft and reported a possible detection.
The first unambiguous observation of p-modes in this star was recently made
with the spectrograph CORALIE mounted on the 1.2-m Swiss telescope at La Silla
Observatory (Bouchy \& Carrier \cite{BC01} (hereafter referred as BC01);
Carrier et al. \cite{carrier01c}). We present here a revised and
extended analysis of the acoustic spectrum of this star and compare it
with theoretical models.

\section{Observations and data reduction}

The observations of $\alpha$~Cen~A were carried out with the CORALIE
fiber-fed echelle spectrograph mounted on the 1.2-m Swiss telescope
at the ESO La Silla Observatory. A description of the spectrograph
and the data reduction process is presented in BC01.

$\alpha$~Cen~A was observed over 13 nights in May 2001. A journal of these
observations is given in Table~1. A total of 1850 optical spectra was 
collected with a typical signal-to-noise ratio in the range of 300\,-\,420 
at 550 nm. Radial velocities were computed for each night relative to the
highest-signal-to-noise-ratio optical spectrum obtained in the middle
of the night (when the target had the highest elevation).
The obtained velocities are shown in Fig.~1. The dispersion
of these measurements reaches 1.53 {\ms} and the individual value for each
night is listed in Table~1.

\begin{figure}
\resizebox{\hsize}{!}{\includegraphics{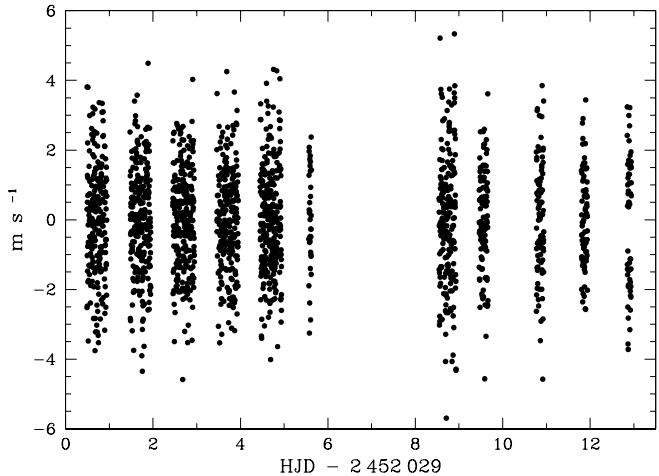}}
\caption{Radial velocity measurements of $\alpha$~Cen~A. A 3 order
polynomial fit subtraction was applied for each night. The dispersion
reaches 1.53 \ms.}
\end{figure}

\begin{table}
\caption{Distribution and dispersion of Doppler measurements}
\begin{center}
\begin{tabular}{clll}
\hline
Date & Nb spectra & Nb hours & $\sigma$ (\ms) \\ \hline
2001/04/29 & 220 & 10.91 & 1.58 \\
2001/04/30 & 259 & 11.36 & 1.49 \\
2001/05/01 & 268 & 11.84 & 1.39 \\
2001/05/02 & 246 & 11.71 & 1.40 \\
2001/05/03 & 267 & 11.77 & 1.53 \\
2001/05/04 &  35 &  1.43 & 1.44 \\
2001/05/05 &  -  &   -   &   -  \\
2001/05/06 &  -  &   -   &   -  \\
2001/05/07 & 206 &  9.32 & 1.84 \\
2001/05/08 & 116 &  4.77 & 1.39 \\
2001/05/09 &  96 &  4.06 & 1.61 \\
2001/05/10 &  81 &  3.30 & 1.34 \\
2001/05/11 &  56 &  2.25 & 1.88 \\ \hline
\end{tabular}
\end{center}
\end{table}

\section{Acoustic spectrum analysis}

\subsection{Power spectrum}

In order to compute the power spectrum of the velocity time series of
Fig.~1,
we used the Lomb-Scargle modified algorithm (Lomb \cite{lomb76}, Scargle
\cite{scargle82}) for unevenly spaced data. The resulting LS periodogram,
shown in Fig.~2, exhibits a series of peaks between 1.8 and 2.9 mHz
modulated
by a broad envelope, which is the typical signature of solar-like
oscillations.
This signature also appears in the power spectrum of each individual night.
Toward the lowest frequencies ($\nu$ $<$ 0.6 mHz), the power rises and
scales inversely with frequency squared as expected for instrumental instabilities. The mean white noise level $\sigma_{\rm ps}$, computed in the range 0.6-1.5 mHz, reaches $2.39\times 10^{-3}$~\m2s2. Considering that this noise is gaussian, the mean noise level in the amplitude spectrum is 4.3 \cms.
With 1850 measurements, the velocity accuracy corresponds thus to 1.05 \ms.

\begin{figure}
\resizebox{\hsize}{!}{\includegraphics{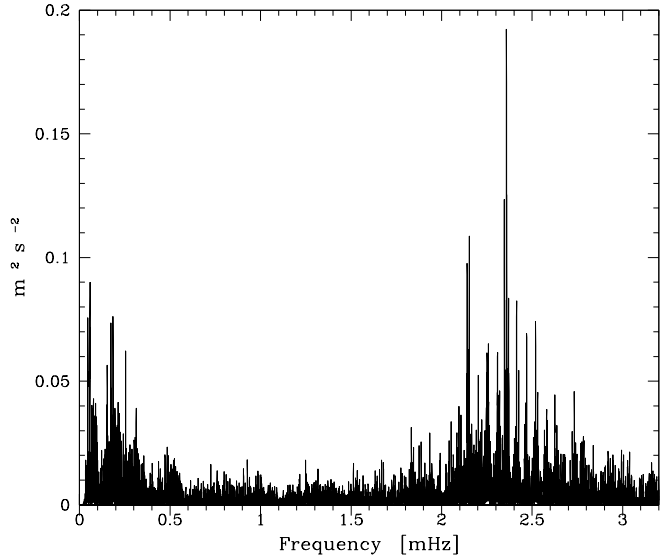}}
\caption{Power spectrum of the radial velocity measurements of
$\alpha$~Cen~A.}
\end{figure}

\subsection{Search for a comb-like pattern}

In solar-like stars, p-mode oscillations are expected to produce a
characteristic comb-like structure in the power spectrum with mode
frequencies
$\nu_{n,l}$ reasonably well approximated by the simplified asymptotic
relation (Tassoul \cite{tassoul80}):
\begin{eqnarray}
\label{eq1}
\nu_{n,l} & \approx &
\Delta\nu_{0}(n+\frac{l}{2}+\epsilon)-l(l+1)\delta\nu_{02}/6
\end{eqnarray}
with $\Delta\nu_{0}\,=\,<\nu_{n,l}-\nu_{n-1,l}>$ and \\
$\delta\nu_{02}\,=\,<\nu_{n,0}-\nu_{n-1,2}>$\,.\\

\begin{figure}
\resizebox{\hsize}{!}{\includegraphics{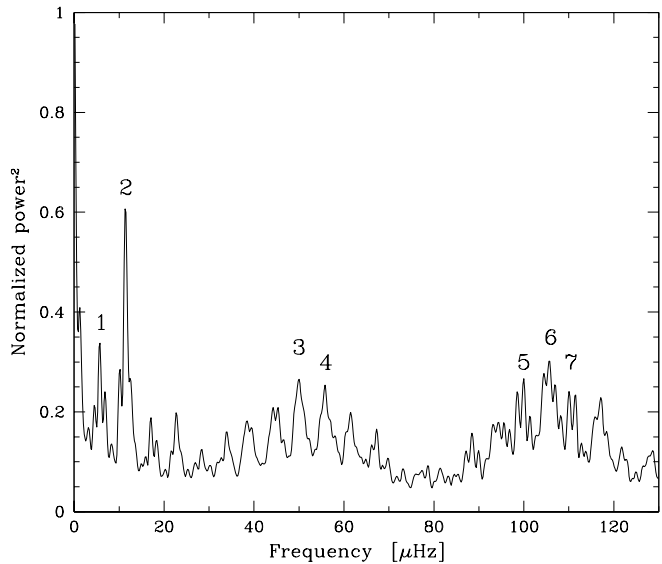}}
\caption{Autocorrelation of the power spectrum of $\alpha$~Cen~A.}
\end{figure}

The two quantum numbers $n$ and $l$ correspond to the radial
order and the angular degree of the modes, respectively. $\Delta\nu_{0}$ and
$\delta\nu_{02}$
are the large and small spacing respectively. For stars of which the disk is
not resolved, only the lowest-degree modes ($l\,\leq\,3$) can be detected.
In case of stellar rotation, high-frequency p-modes need to be
characterized by a third quantum number $m$ called the azimuthal order:
\begin{eqnarray}
\label{eq2}
\nu_{n,l,m} & \approx & \nu_{n,l,0}+m/P_{\rm rot}
\end{eqnarray}
with $-l\,\leq\,m\,\leq\,l$ and $P_{\rm rot}$ the averaged stellar rotational
period.\\

One technique, commonly used to search for periodicity in the power
spectrum,
is to compute its autocorrelation. The result is shown in Fig.~3. The first
identified peak at 5.7~$\mu$Hz corresponds to correlation between
modes $l$=2 and $l$=0, hence the small spacing
$\delta\nu_{02}$. The second peak at 11.5~$\mu$Hz corresponds to the daily
alias. Peak 3 at 50~$\mu$Hz is associated with correlation between
modes $l$=0 and $l$=1 and/or $l$=1 and $l$=2. The fourth peak at
55.8~$\mu$Hz
is associated with correlation between modes $l$=2 and $l$=1 and/or $l$=1
and $l$=0. The separation between peaks 3 and 4 corresponds to the small
spacing $\delta\nu_{02}$. Peaks 5 and 7, at 99.9 and 111.4~$\mu$Hz 
respectively, are associated with correlations between modes $l$=0 and $l$=2.
Finally peak 6 at 105.5~$\mu$Hz corresponds to the large spacing
$\Delta\nu_{0}$. Other peaks correspond to correlations between modes and
daily aliases.

\subsection{Echelle diagram}

In order to identify the degree $l$ of each mode individually,
the power spectrum has been cut into slices of 105.5 $\mu$Hz which are 
displayed above one another. Such a process was proposed and used by
Grec et al. (\cite{grec83}) to identify the degree $l$ of solar modes. 
The result, also called echelle diagram, is presented in Fig.~4.

\begin{figure}
\resizebox{\hsize}{!}{\includegraphics{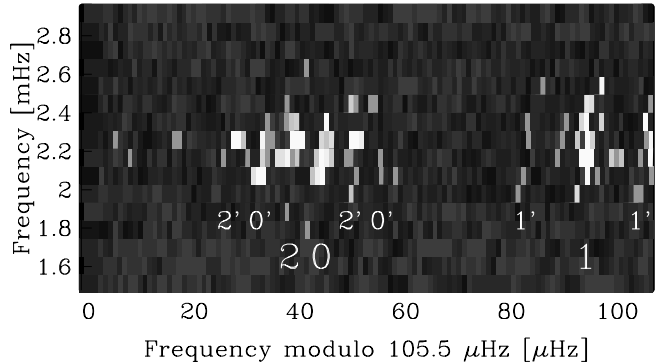}}
\caption{Echelle diagram of $\alpha$ Cen A oscillations. Modes $l$ = 0, 1
and 2
and their aliases are identified.}
\end{figure}

Fig.~5 corresponds to the merged echelle diagram, all the slices
of 105.5 $\mu$Hz were summed up.

Both Figs.~4 and 5 make it possible to identify modes of degree $l$=0, 1 and
2 and their side-lobes due to the daily aliases at $\pm$ 11.57 $\mu$Hz.
These two figures indicate that the modes follow an asymptotic
relation reasonably well.

\begin{figure}
\resizebox{\hsize}{!}{\includegraphics{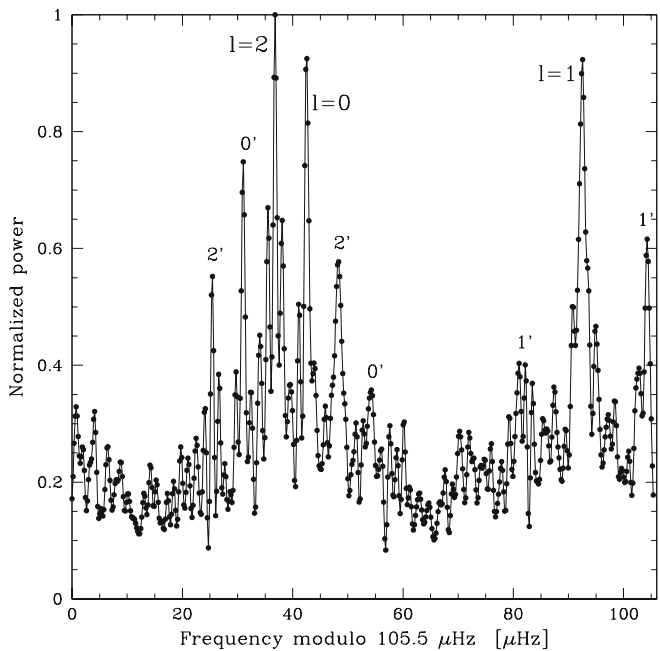}}
\caption{Sum of the echelle diagram of $\alpha$ Cen A oscillations. Modes
$l$ = 0, 1 and 2 and their aliases are identified.}
\end{figure}

\subsection{Gap filling process}

In order to reduce the effect of single-site observation, the
repetitive music process proposed by Fossat et al. (\cite{fossat99}) was
used. This method is based on the fact that the autocorrelation function of the
full disk helioseismological signal, after dropping quickly to zero in 20 or 30
minutes, shows secondary quasi periodic bumps, due to the quasi-periodicity
of the peak distribution in the Fourier spectrum. In the helioseismological
signal, the first of these bumps appears at about 4.1 hours and corresponds
to a $\Delta\nu_{0}/2$ = 67.5 $\mu$Hz periodicity of the modes. An obvious
consequence is that simply replacing a gap by the signal collected 4.1 hours 
earlier or 4.1 hours later provides a very efficient gap filling method.
It should be noted that such a process modulates the background noise with 
the 67.5 $\mu$Hz periodicity, the fine adjustment of the method consists of
placing the maxima of this modulation on the p-mode frequencies, so that 
any loss of information is located in the noise.
This gap-filling process is not suitable if the modes do not follow
an asymptotic relation. As precaution the frequency range should be cut 
in several parts and the fine tuning of the temporal periodicity
adjusted in each of them.

In the case of $\alpha$ Cen A, the periodicity of the modes is equal to
$\Delta\nu_{0}/2$ = 52.8 $\mu$Hz which correspond to a periodicity in time
of the signal of about 5.26 hours. Considering that $\alpha$ Cen A was 
observed during the longest nights for a little less than 12 hours, this gap-filling process makes it possible to fill the duty cycle up to 90 \%.
The power spectrum of a pure sinusoid using the observational window of
$\alpha$~Cen~A with and without gap filling is presented in Fig.~6.
Note that aliases at $\pm$ 1.35 $\mu$Hz are not reduced, due to the gap of
the seventh and eighth nights during the run.

\begin{figure}
\resizebox{\hsize}{!}{\includegraphics{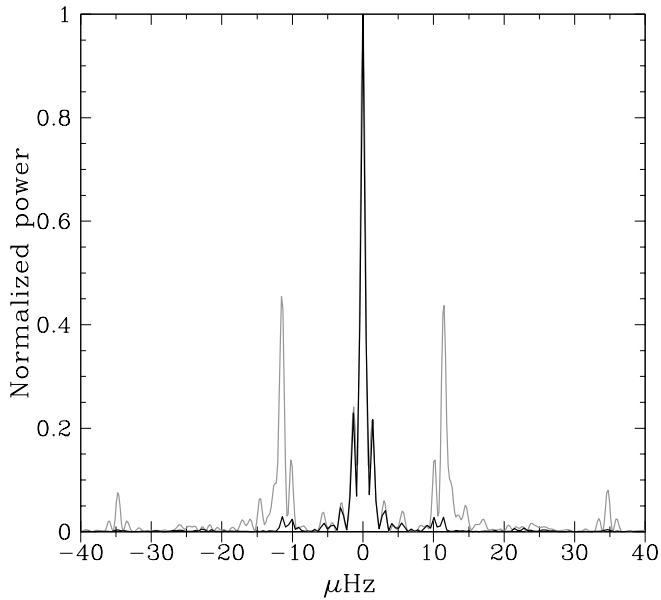}}
\caption{Observational window response with (black line)
and without (grey line) the gap-filling process.}
\end{figure}

The frequency range was cut into 3 parts (1.8 - 2.2 mHz, 2.15 - 2.55 mHz
and 2.5 - 2.9 mHz) and the temporal periodicity was adjusted in each of them 
to place the maxima of the modulation on modes $l$=1 which are easily
identifiable. For the 3 parts of the power spectrum, the temporal
periodicity used was respectively 5.299, 5.294 and 5.289 hours. 
The result is presented in Fig.~7.

This gap-filling process does not cause a drastic change in the power
spectrum below 2.5 mHz. Above 2.5 mHz, the process helps in the identification
of modes, especially modes $l$=0 and $l$=2 by removing or reducing aliases.
Note that this process is based on the assumption that frequencies follow
an asymptotic relation and can destroy any other frequency 
and/or enhance their side-lobes.

\begin{figure*}
\resizebox{\hsize}{!}{\includegraphics{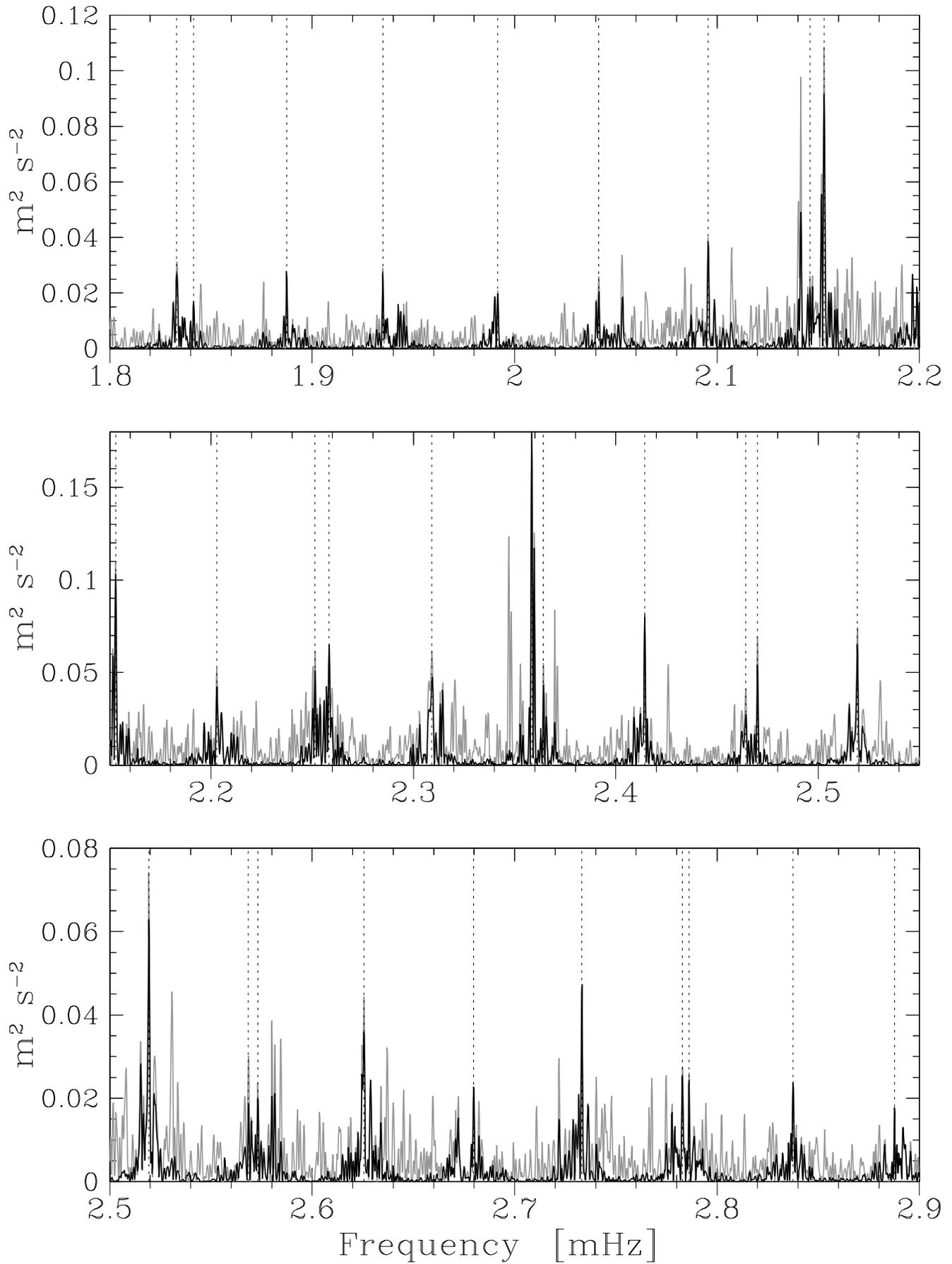}}
\caption{Power spectrum of $\alpha$~Cen~A with (black line) and without
(grey line) the gap filling process. The identified frequencies are indicated with dotted lines.}
\end{figure*}

\subsection{Frequencies of p-modes}

The strongest peaks of the power spectrum computed with the gap filling
process were identified and are listed in Table~2. These
peaks were selected such as to have a signal-to-noise ratio in the 
amplitude spectrum greater than 3 (0.017~\m2s2 in the power spectrum).

The frequency resolution of our time series is 0.93 $\mu$Hz. If we suppose
that no mode is resolved, the estimated uncertainty on the
frequency determination is 0.46 $\mu$Hz. The values of $n$ and $l$
are deduced from the asymptotic relation (see Eq.~(1)) assuming that the
parameter $\epsilon$ is near the solar value ($\epsilon_{\odot} \sim 1.5$).
Our first identification in
BC01 based on the 5 first nights did not make a distinction between the
modes $l$=0 and $l$=2. The last row in Table~2 gives the average large
spacing for each angular degree. All the identified modes in Table~2 are
indicated in Fig.~7.

\begin{table}
\caption{Mode frequencies (in $\mu$Hz) of $\alpha$~Cen~A. Uncertainty on
frequency is estimated to 0.46 $\mu$Hz.}
\begin{center}
\begin{tabular}{lccc}
\hline
 & $l$ = 0 & $l$ = 1 & $l$ = 2 \\
\hline
n = 15 &         &         & 1833.1 \\
n = 16 & 1841.3  & 1887.4  & 1934.9 \\
n = 17 &         & 1991.7  & 2041.5 \\
n = 18 &         & 2095.6  & 2146.0 \\
n = 19 & 2152.9  & 2202.8  & 2251.4 \\
n = 20 & 2258.4  & 2309.1  & 2358.4 \\
n = 21 & 2364.2  & 2414.3  & 2464.1 \\
n = 22 & 2470.0  & 2519.3  & 2568.5 \\
n = 23 & 2573.1  & 2625.6  &        \\
n = 24 & 2679.8  & 2733.2  & 2782.9 \\
n = 25 & 2786.2  & 2837.6  & 2887.7 \\
\hline
$\Delta\nu_0$ & 105.6 $\pm$ 0.5 & 105.6 $\pm$ 0.4 & 105.0 $\pm$ 0.5 \\
\hline
\end{tabular}\\
\end{center}
\end{table}

The average large and small spacing and the parameter $\epsilon$ are 
deduced from a least-squares fit of Eq.~(1) with the frequencies of Table~2:

\begin{displaymath}
\Delta\nu_{0}\,=\,105.5\,\pm\,0.1\;\mu Hz,
\end{displaymath}
\begin{displaymath}
\delta\nu_{02}\,=\,5.6\,\pm\,0.7\;\mu Hz,
\end{displaymath}
\begin{displaymath}
\epsilon\,=\,1.40\,\pm\,0.02\;.
\end{displaymath}

Fig.~8 represents the echelle diagram of peaks listed in Table~2.

\begin{figure}
\resizebox{\hsize}{!}{\includegraphics{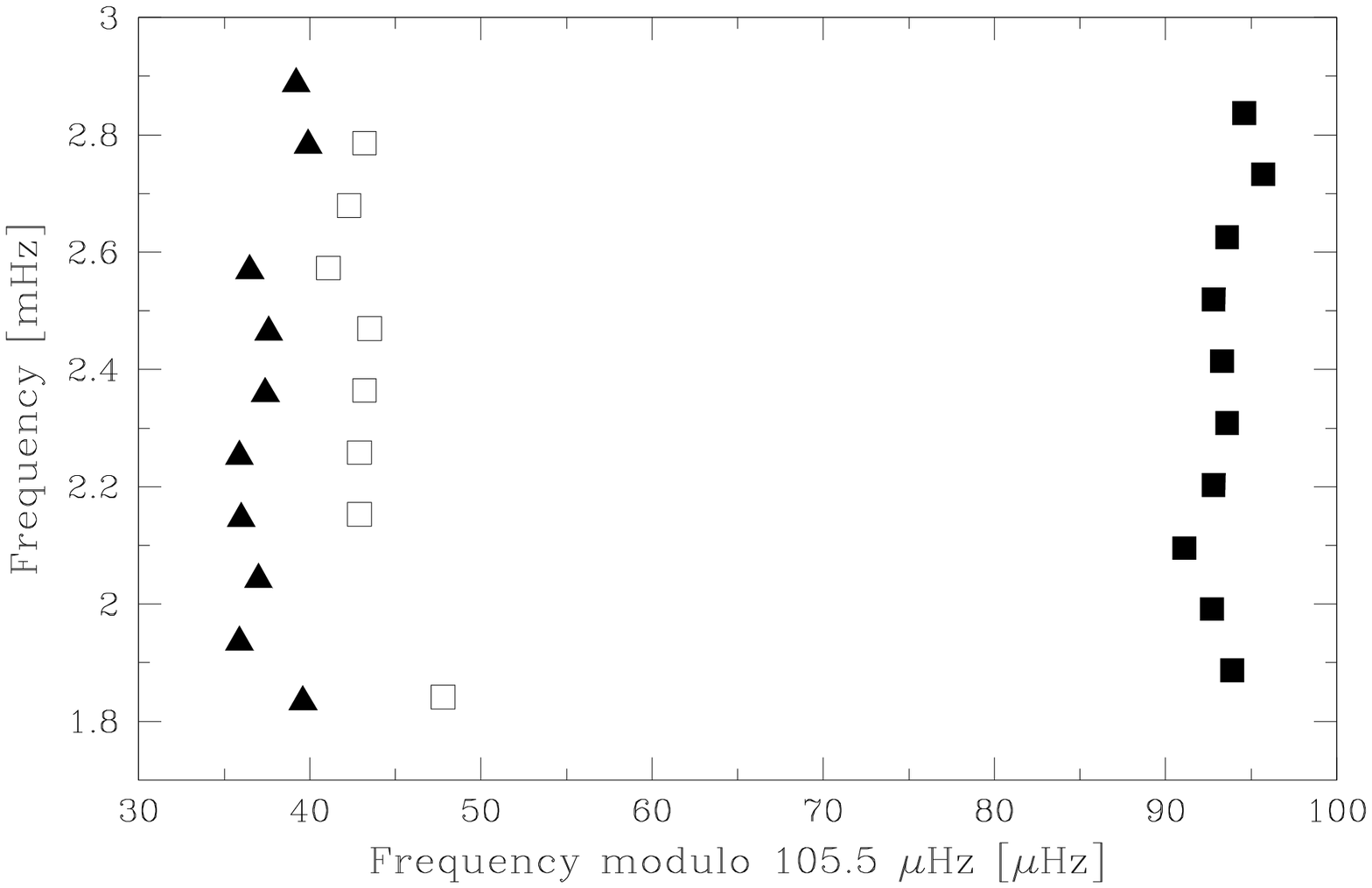}}
\caption{Echelle diagram of identified modes in the power spectrum. The
modes $l$=2 ($\blacktriangle$), $l$=0 ($\square$) and $l$=1 ($\blacksquare$) 
are represented from left to right.}
\end{figure}

\subsection{Amplitudes of p-modes}

In order to estimate the average amplitude of the modes, we followed the
method of Kjeldsen \& Bedding (\cite{kjeldsen95}). A simulated
time series was generated with artificial signal plus noise using our window function. The generated signal corresponds to modes centered on 2.36 mHz,
separated by $\Delta\nu_{0}/2$ and convolved by a gaussian envelope. The
generated noise is gaussian with a standard deviation of 1 \ms. Several simulations show that the envelope amplitude is in the range 29\,-\,33 \cms.

Table~3 and Fig.~9 present the estimated Doppler amplitude of each
identified mode with the assumption that none of them is resolved. This
amplitude was determined as the height of the peak in the power spectrum
after quadratic subtraction of the mean noise level.
Fig.~9, like Fig.~5, shows that the modes $l$=0, 1 and 2 have a
similar average amplitude.

\begin{table}
\caption{Doppler amplitude of identified modes determined as the height of
the peak in the power spectrum after quadratic subtraction of the mean noise
level.}
\begin{center}
\begin{tabular}{llcc}
\hline
 $n$ & $l$ & $\nu_{n,l}$ & A$_{n,l}$  \\
   &   & $\mu$Hz     & {\cms}     \\
\hline
 15 & 2 & 1833.1 & 17 \\
 16 & 0 & 1841.3 & 12 \\
 16 & 1 & 1887.4 & 15 \\
 16 & 2 & 1934.9 & 16 \\
 17 & 1 & 1991.7 & 14 \\
 17 & 2 & 2041.5 & 15 \\
 18 & 1 & 2095.6 & 19 \\
 18 & 2 & 2146.0 & 15 \\
 19 & 0 & 2152.9 & 33 \\
 19 & 1 & 2202.8 & 22 \\
 19 & 2 & 2251.4 & 24 \\
 20 & 0 & 2258.4 & 25 \\
 20 & 1 & 2309.1 & 24 \\
 20 & 2 & 2358.4 & 44 \\
 21 & 0 & 2364.2 & 23 \\
 21 & 1 & 2414.3 & 28 \\
 21 & 2 & 2464.1 & 20 \\
 22 & 0 & 2470.0 & 26 \\
 22 & 1 & 2519.3 & 27 \\
 22 & 2 & 2568.5 & 17 \\
 23 & 0 & 2573.1 & 15 \\
 23 & 1 & 2625.6 & 21 \\
 24 & 0 & 2679.8 & 13 \\
 24 & 1 & 2733.2 & 21 \\
 24 & 2 & 2782.9 & 16 \\
 25 & 0 & 2786.2 & 15 \\
 25 & 1 & 2837.6 & 15 \\
 25 & 2 & 2887.7 & 12 \\
\hline
\end{tabular}
\end{center}
\end{table}

\begin{figure}
\resizebox{\hsize}{!}{\includegraphics{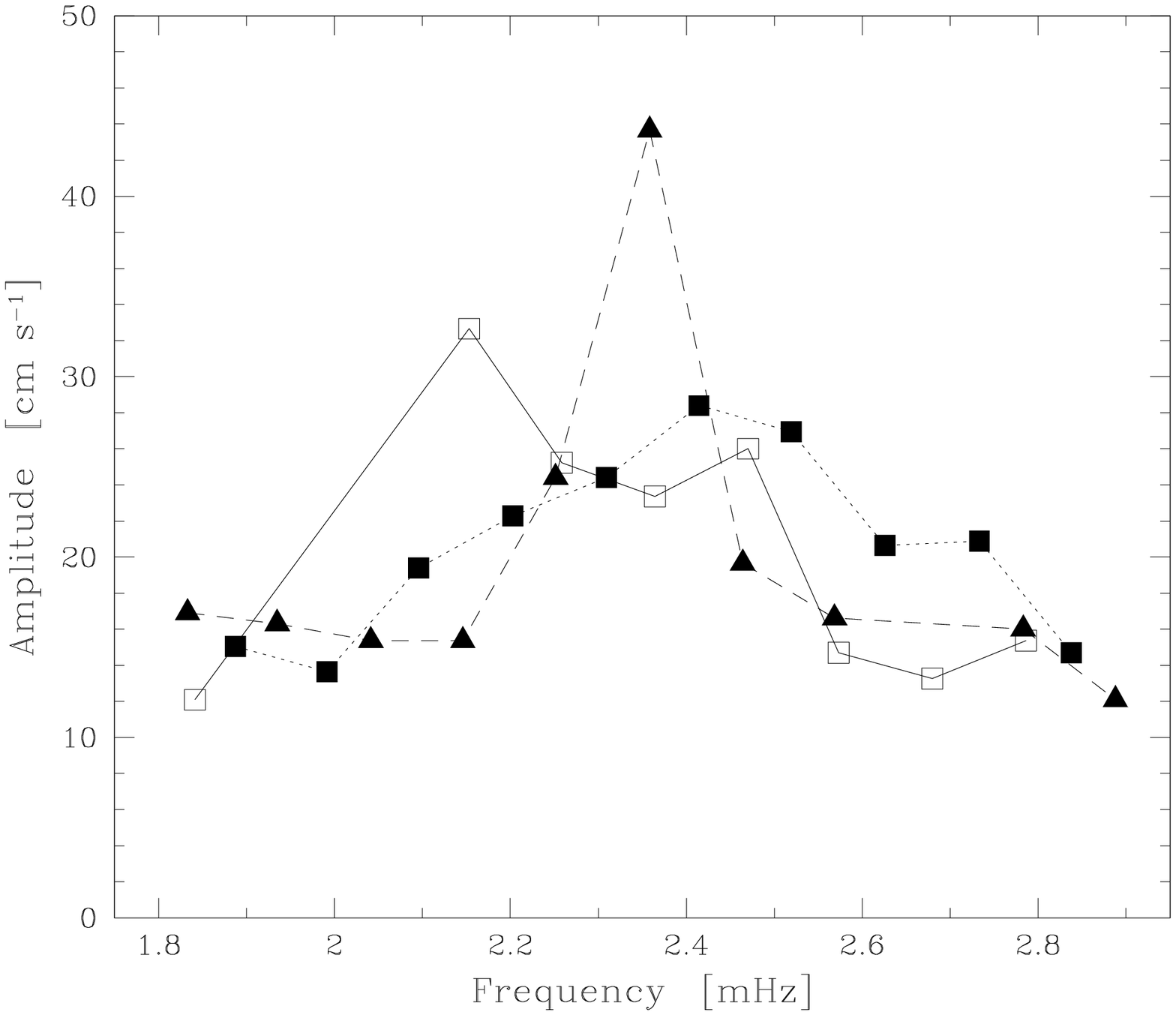}}
\caption{Doppler amplitude of identified modes of degree $l$=0 ($\square$),
$l$=1 ($\blacksquare$) and $l$=2 ($\blacktriangle$). }
\end{figure}

In order to compare the observed amplitude with the expected one,
we computed the amplitude ratios between mode $l$=0 and the other
modes $(l,m)$. These amplitude ratios (see Table~4) were determined
following Gouttebroze \& Toutain (\cite{goutte94}) assuming that all
the modes have the same intrinsic amplitude. The effect of limb darkening was neglected and the inclination of the rotational axis was assumed equal to the inclination of the orbital axis of the binary system ($79^o$) determined by Pourbaix et al. (\cite{pourbaix99}).

\begin{table}
\caption{Expected amplitude ratios of modes.}
\begin{center}
\begin{tabular}{lc}
\hline
 $l$,$m$ & amplitude ratios  \\
\hline
$l$=0,$m$=0     & 1.00  \\
$l$=1,$m$=0     & 0.25  \\
$l$=1,$m$=-1,1  & 0.90  \\
$l$=2,$m$=0     & 0.40  \\
$l$=2,$m$=-1,1  & 0.21  \\
$l$=2,$m$=-2,2  & 0.53  \\
$l$=3,$m$=0     & 0.09  \\
$l$=3,$m$=-1,1  & 0.12  \\
$l$=3,$m$=-2,2  & 0.08  \\
$l$=3,$m$=-3,3  & 0.18  \\
\hline
\end{tabular}
\end{center}
\end{table}

The observed amplitude of modes $l$=2 is higher than expected; this 
can be due to interferences between modes of different $m$.
The modes corresponding to $l$=3 are expected to have an amplitude of only
20 \% of the $l$=1 mode, hence to be well below the threshold of
identification.

\subsection{Linewidth of p-modes and rotational splitting}

The expected linewidth for a 1.1 M$_{\odot}$ star in the frequency range 
1.8 to 2.9 mHz is less than 2 $\mu$Hz (Houdek \cite{houdek99}).
Our frequency resolution of 0.93 $\mu$Hz is then too low to attempt
a clear determination of the linewidth of the identified modes.
Below 2.2 mHz, modes seem unresolved and present a linewidth
equal to the frequency resolution, hence a damping time longer than 25
days. Above 2.2 mHz, the structure of modes seems more complex. This could 
be eventually caused by the fact that their linewidths begin to be greater than 
the frequency resolution.

For $\alpha$ Cen A, Hallam et al. (\cite{hallam91}) estimated a rotational
period of 28.8 $\pm$ 2.5 days. More recently Saar \& Osten (\cite{saar97})
measured a rotational velocity of 2.7 $\pm$ 0.7~\kms. With an estimated
radius of 1.2 R$_{\odot}$, the period of rotation is 22 days.
Assuming a uniform rotation, the corresponding splitting of the
modes is expected to be about 0.5 $\mu$Hz. Assuming an inclination of the
rotational axis of 79$^o$, a separation of about 1 $\mu$Hz should be visible 
between modes ($l$=1,$m$=-1) and ($l$=1,$m$=1), and modes ($l$=2,$m$=0) and
($l$=2,$m$=$\pm$2).

\begin{figure}
\resizebox{\hsize}{!}{\includegraphics{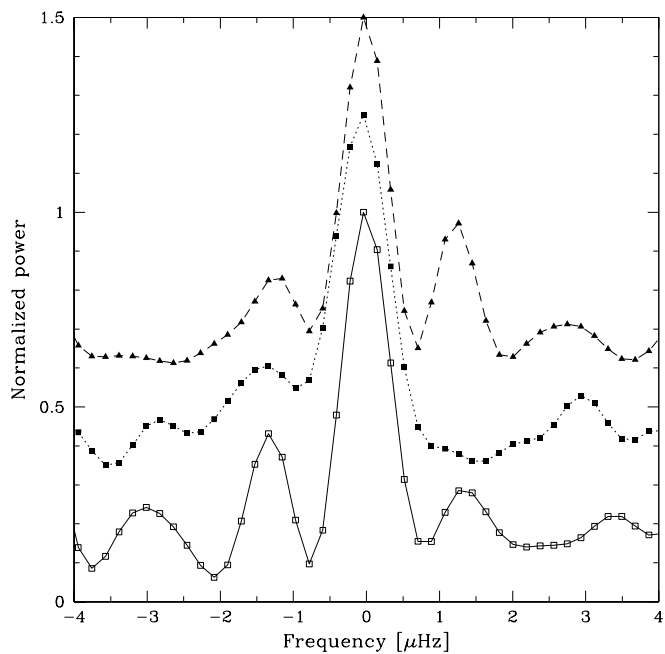}}
\caption{Superposition of identified modes $l$=0 ($\square$), $l$=1
($\blacksquare$) and $l$=2 ($\blacktriangle$) in the power spectrum with
frequency shifted to zero.}
\end{figure}

Figure~10 shows the superposition in the power spectrum of modes $l$=0,
$l$=1 and $l$=2 with the frequency of each mode shifted to
zero. Modes $l$=0 present side lobes at $\pm$~1.35~$\mu$Hz, probably
introduced by the aliases due to the gap of the seventh and eighth nights 
during the run.
Modes $l$=2 present side lobes at $\pm$ 1.25 $\mu$Hz which could be
introduced by these same aliases or/and a rotational splitting. In the 
latter case it should correspond to a period of rotation of about 19 days, 
lower than the expected one. Modes $l$=1, which show a side lobe only on the left, do not allow a conclusion. The CLEAN algorithm (Roberts et al. \cite{roberts87}) was used in order to remove the effect of the observational window response.  We noticed that this process does not eliminate the whole signal at modes $l$=2, as is the case for $l$=0 and $l$=1.
The pattern shown in Fig.~10 leads one to think that an error of
$\pm$~1.3~$\mu$Hz could have been introduced at some identified mode 
frequencies and could explain the
dispersion of the mode frequencies around the asymptotic relation shown in
Fig.~8.

Additional data will be needed to determine both the rotational splitting 
and the damping time of $\alpha$~Cen~A p-modes.

\section{Comparison with models}

Seismological parameters deduced from our observations are in full 
agreement with the expected values scaling from the Sun (Kjeldsen \& Bedding,
\cite{kjeldsen95}) giving the frequency of the greatest mode
$\nu_{\rm max}\,=\,2.3$ mHz, the large spacing
$\Delta\nu_{\rm 0}\,=\,105.8$~$\mu$Hz and the oscillation peak amplitude
$A_{\rm osc}\,=\,31.1$ {\cms}.

The variations with frequency of the large and small spacings are
compared in Figs.~11 and 12 with the values deduced from
models recently developed by Morel et al. (\cite{morel00}) and Guenther \&
Demarque (\cite{guenther00}).
The observational constraints for the model calibrations and
the properties of the calibrated models are given in Table~5. The two sets
of models differ essentially by the estimated mass and the age of the star.
The large spacing, mainly related to stellar density, varies proportional 
to $M^{1/2}\,R^{-3/2}$. The small spacing, mainly related to the structure of
the core, decreases with stellar age and mass.

Our observations are well bracketed by the two sets of models for both the
large and small spacing. This clearly suggests a model with intermediate properties, i.e. a mass between 1.10 and 1.16 M$_{\odot}$.
New models of $\alpha$~Cen~A are being developed by Thevenin et al.
(\cite{thevenin02}), with different properties in order to better interpret
the observed oscillations and constraint the physical parameters
of this star.\\

\begin{table}
\caption{Observational constraints and global characteristics of the
models.}
\begin{center}
\begin{tabular}{lcc}
\hline
Models & Morel et al. &  Guenther \& Demarque \\
\hline
$M/M_{\odot}$ &  $1.16\pm0.031$     &  $1.1015\pm0.008$ \\
$T_{\rm eff} [{\rm K}]$ &  $5790\pm30   $     &  $5770\pm50 $\\
$\log g$      &  $4.32\pm0.05 $     &  $4.28\pm0.02$ \\
$\rm [\frac{Fe}H]_i$    &  $0.20\pm0.02 $     &  $0.22\pm0.02$ \\
$L/L_{\odot}$ &  $1.534\pm0.103 $   &  $1.572\pm0.135 $\\
\hline
$t [{\rm Myr}]$       & 2710   & 5640   \\
$Y_{\rm i}$     & 0.284  & 0.300 \\
$(\frac ZX)_{\rm i}$ & 0.0443 & 0.0480\\
$\alpha$  & 1.53  & 1.86  \\
\hline
\end{tabular}\\
\end{center}
\end{table}

\begin{figure}
\resizebox{\hsize}{!}{\includegraphics{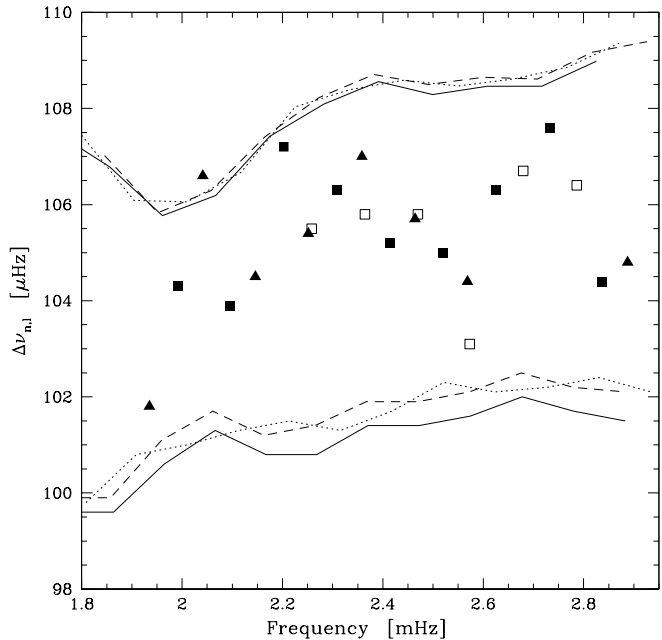}}
\caption{Variations of the large spacing between modes with
consecutive radial order $\Delta\nu_{n,l}=\nu_{n,l}-\nu_{n-1,l}$ for
p-modes of degree $l$=0 ($\square$), $l$=1 ($\blacksquare$) and $l$=2
($\blacktriangle$) versus frequency. Upper lines correspond to the model
of Morel et al. and lower lines to that of Guenther \& Demarque.
Solid, dashed and dotted lines correspond respectively to modes $l$=0, 1 and
2.}
\end{figure}

\begin{figure}
\resizebox{\hsize}{!}{\includegraphics{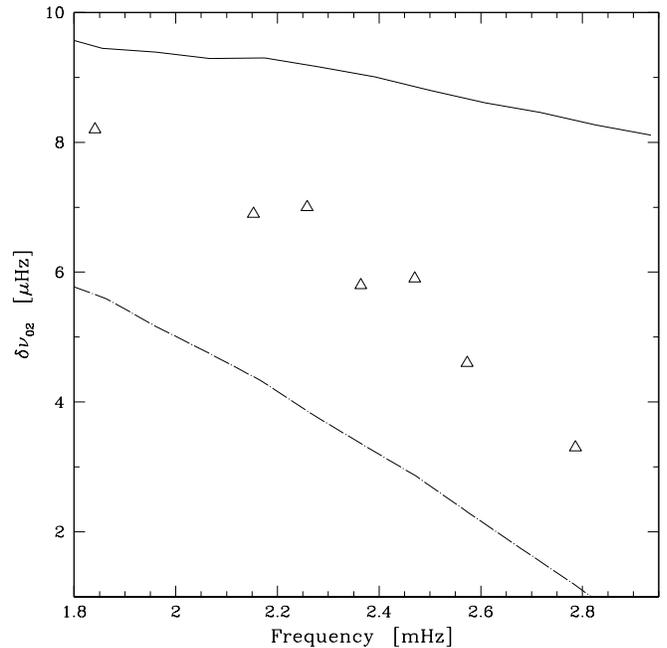}}
\caption{Variations of the small frequency spacing
$\delta\nu_{02}= \nu_{n,0}-\nu_{n-1,2}$ versus frequency. The upper line
corresponds to the model of Morel et al. and the lower line to that of
Guenther \& Demarque.}
\end{figure}

\section{Conclusions and prospects}

Our observations of $\alpha$~Cen~A yield a clear detection of p-mode
oscillations. Several identifiable modes appear in the power spectrum
between 1.8 and 2.9 mHz with an average large spacing of 105.5 $\mu$Hz, an average small spacing of 5.6 $\mu$Hz and an envelope amplitude of about 
31 \cms. These characteristics, in full agreement with the expected
values, make it possible to put constraints on the physical parameters of
this star.

Additional data with a higher signal-to-noise
and a higher frequency resolution will make it possible to determine
properly the suspected rotational splitting and the damping time
of $\alpha$~Cen~A p-modes.

This result, obtained with a small telescope, demonstrates the power of
Doppler ground-based asteroseismology. Future spectrographs like HARPS
(Queloz et al. \cite{queloz01}; Bouchy \cite{bouchy01}) are expected to
conduct asteroseismological study on a large sample of solar-like
stars, and to enlarge significantly our understanding of stellar physics.

\begin{acknowledgements}
We are very obliged to M. Mayor who encouraged our program and allocated 
us time on the Euler Swiss telescope. D. Queloz and L. Weber are acknowledged
for their help in the adaptation of the reduction pipeline needed for our
observation sequences. The authors wish to thank Hans Kjeldsen for offering
valuable suggestions and stimulating discussions.
This work was financially supported by the Swiss National Science
Foundation.
\end{acknowledgements}

\end{document}